\documentclass[12pt,preprint]{aastex}

\usepackage{graphicx}
\usepackage{rotating}
\shorttitle{}
\shortauthors{Nesvorn\'y et al.}

\begin{document}
\baselineskip 19.pt

\title{A Spiral Structure in the Inner Oort Cloud}

\author{David Nesvorn\'y$^1$, Luke Dones$^1$, David Vokrouhlick\'y$^2$, Hal F. Levison$^1$, Cristian Beaug\'e$^3$,
Jacqueline Faherty$^4$, Carter Emmart$^4$, Jon P. Parker$^4$}

\affil{(1) Department of Space Studies, Southwest Research Institute, 1301 Walnut St., 
Suite 400,  Boulder, CO 80302, USA}

\affil{(2) Institute of Astronomy, Charles University, V Hole\v{s}ovi\v{c}k\'ach 2, 
  CZ–18000 Prague 8, Czech Republic}

\affil{(3) Instituto de Astronom\'{\i}a Te\'orica y Experimental (IATE), Observatorio Astron\'omico,
Universidad Nacional de C\'ordoba, Laprida 854, X5000BGR C\'ordoba, Argentina}

\affil{(4) Department of Astrophysics, American Museum of Natural History, 200 Central Park W, New York,
  NY 10024, USA}

\begin{abstract}
As the Galactic tide acts to decouple bodies from the scattered disk it creates a spiral structure 
in physical space that is roughly 15,000 au in length. The spiral is long-lived and persists 
in the inner Oort cloud to the present time. Here we discuss dynamics underlying the Oort spiral 
and (feeble) prospects for its observational detection.  
\end{abstract}

\section{Introduction}
The Oort cloud is a large shell of icy bodies surrounding the solar system at heliocentric 
distances $1,\!000 \lesssim r \lesssim 100,\!000$ au. These bodies are faint and not directly 
observed but their existence is inferred from observations of long-period comets (LPCs; 
Oort 1950). The so-called {\it new} LPCs, which are observed during their first perihelion 
passage through the inner Solar System, often have the semimajor axes between $20,\!000$ and 
$100,\!000$ au.\footnote{According to Kr\'olikowska \& Dybczy\'nski (2017), comets must have
  original semimajor axes of at least 20,000 au to be new. Comets with $20,\!000<a<40,\!000$ au
  are a mix of new and old. Comets with $a>40,\!000$ au are new.}
They are thought to have relatively recently evolved, due to the effects of the 
Galactic tide (Heisler \& Tremaine 1986; Section 3 here), onto high-eccentricity/low-perihelion 
orbits. The new LPCs have a nearly isotropic distribution of orbital inclinations suggesting 
that the outer Oort cloud at $r > 10,000$ au is roughly spherical (see Dones et al. 2015 for
a review). 

Oort cloud formation dates back to early stages of the solar system some 4.6 Gyr 
ago (Duncan et al. 1987, Dones et al. 2004, Vokrouhlick\'y et al. 2019). First, as the outer planets
cleared their orbital neighborhood, small bodies were scattered onto very eccentric orbits
with perihelion distances $q \lesssim 
30$ au and semimajor axes $a \gtrsim 1,\!000$ au (orbital eccentricities $e \gtrsim 0.97$). 
Second, the Galactic tide raised the perihelion distances of these bodies, effectively decoupling 
them from planetary perturbations, and tilted their orbits. 
Third, encounters of the Sun with stars in the Galactic neighborhood thoroughly 
mixed the orbits in the outer Oort cloud, producing a relatively homogeneous and isotropic 
source for LPCs.\footnote{Most bodies from $r<30$ au were ejected to interstellar space -- only 
a relatively small fraction ended up in the Oort cloud.}

Dynamical simulations reveal formation of the {\it inner} Oort cloud at $1,\!000 < r < 10,\!000$~au
(Duncan et al. 1987, Levison et al. 2001, Vokrouhlick\'y et al. 2019).\footnote{
  The inner edge location of the cloud depends on the strength of the Galactic tide. It shifts inward
  if the Sun formed closer to the Galactic center and migrated out (Brasser et al. 2010, Kaib et al. 2011).
  In addition, a massive inner Oort cloud extending down to $\sim 300$--500 au would form if/when the Sun
  resided in its birth stellar cluster (Nesvorn\'y et al. 2023 and references therein).}
The inner Oort cloud forms in much the same way as the outer Oort cloud, except that the 
timescale on which the Galactic tide changes orbits at $1,\!000 < r < 10,\!000$ au is long, comparable
to the age of the Solar System. This explains why the inner Oort cloud is not a dominant source 
of LPCs (Vokrouhlick\'y et al. 2019): bodies from this region evolve too slowly and are ejected by planets before they 
can reach $q \lesssim 3$ au, heat up and become active comets (Hills 1981; but see Kaib \& Quinn
2009).\footnote{The population of new LPCs
with $1,\!000 < a < 10,\!000$ may become more noticeable as observations start characterizing 
the LPC population with larger perihelion distances ($q>5$ au; Vokrouhlick\'y et al. 2019). The inner Oort
cloud may be activated and produce more LPCs immediately after a significant stellar encounter
(Vokrouhlick\'y et al. 2019).} In addition, the orbits with 
$1,\!000 < a < 10,\!000$ au, which are more strongly bound to the Sun, are less affected by stellar 
encounters. The inner Oort cloud is therefore often portrayed as a relatively flat disk, roughly 
aligned with the ecliptic (Levison et al. 2001), that retained memory of its initial conditions 
(e.g., Fouchard et al. 2018).

In the inner Oort cloud at $1,\!000 < r < 10,\!000$ au, structures in the spatial distribution
of bodies can form and `freeze' over timescales comparable to the age of the Solar System. This raises 
the question of how the inner Oort cloud would look to a distant observer and/or whether there are 
any diagnostic features that would facilitate its detection. Here we show that the inner 
Oort cloud is a slightly warped disk, roughly 15,000 au across, inclined $i \sim 30^\circ$ to the 
ecliptic (nearly polar in the Galactic reference system, Galactic inclination $i_{\rm G} \sim 
90^\circ$). The disk, when viewed from distance, would appear as a spiral structure with two twisted 
arms. In section 2, we discuss the results of dynamical simulations -- described in Appendix A -- 
to illustrate the inner Oort cloud structure. The analytic model of Breiter \& Ratajczak (2005) (also 
see Higuchi et al. 2007) is employed to interpret these results (Section 3). Observational 
detectability is discussed in Section 4.

\section{The Oort spiral}
 
\begin{figure}
\epsscale{0.7}
%gr.faherty2.f > faherty2.dat in Cluster/galaxy, and anim.f 
\plotone{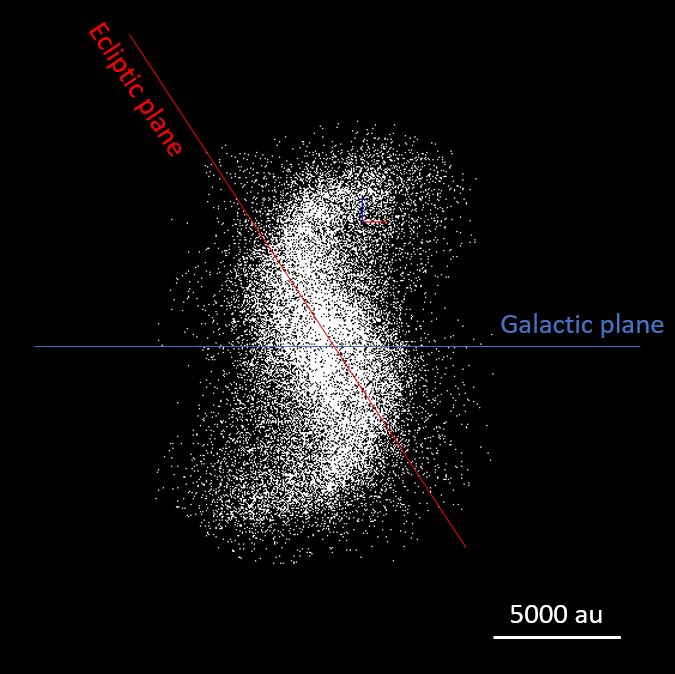}
\caption{The spiral structure in the inner Oort cloud viewed by a distant observer along the Galactic node 
direction (intersection of the Galactic and ecliptic planes). The distribution of bodies was obtained from
the {\it Galaxy} simulation in Nesvorn\'y et al. (2023). To show things clearly, here we isolated the inner 
Oort cloud from the more spherical outer Oort cloud by plotting bodies, 34,000 in total, with $a<5,\!000$ au. 
If the {\it outer} Oort cloud were plotted in the figure, it would appear as a large, roughly spherical 
envelope of the spiral. The classical Kuiper belt with $r<100$~au is not shown here -- it would appear as a 
doughnut-like central concentration of bodies aligned with the ecliptic plane.}           
\label{spiral}
\end{figure}

Figure \ref{spiral} shows the Oort spiral as it should appear at the present epoch. The distribution
of bodies in the inner Oort cloud was extracted from the {\it Galaxy} simulation (Nesvorn\'y et al. 2023;
see Appendix A here for a description of the simulation setup)
and was plotted from a viewpoint of a distant observer. The observer is located at the intersection
of the Galactic and ecliptic planes, with the Galactic plane running horizontally across the plot, and
the ecliptic plane  
tilted 60.2 deg to it along the main axis of the spiral. The Sun is in the plot's center. The main axis 
of the spiral is aligned with the ecliptic -- the ends of the spiral are twisted away from the ecliptic.   
We note that Figure \ref{spiral} does not incorporate any information about the actual detectability of 
the inner Oort cloud (for that, see Section 4).       

We verified that the spiral exists in all our previous simulations with the Galactic tide independently 
of whether the effects of the stellar cluster were included (Nesvorn\'y et al. 2017, 2023; Vokrouhlick\'y 
et al. 2019). The spiral is long-lived: it emerges in the first hundreds of Myr after the formation of 
the solar system and persists over billions of years. The same spiral structure appears in simulations 
with different sequences of stellar encounters, indicating that the spiral is not related to stellar 
encounters. Instead, as we establish below, the spiral is a consequence of the Galactic tide. 
To simulate the Galactic tide, the {\it Galaxy} simulation in Nesvorn\'y et al. (2023) adopted the 
mass density of 0.15 M$_{\rm Sun}$ pc$^{-3}$ in the solar neighborhood. By comparing the results of simulations with 
different mass densities, we found that the spiral becomes smaller (larger) for higher (lower) 
stellar densities in the solar neighborhood. This suggests a direct involvement of the Galactic tide.     

We studied formation of the Oort spiral by inspecting orbital histories of bodies contributing to the spiral.
They were first scattered to $1,\!000<a<10,\!000$ au by migrating planets (Neptune, Uranus and Saturn).
The scattered orbits had relatively low orbital inclinations with respect to the ecliptic 
($i \lesssim 30^\circ$), low perihelion distances ($q \lesssim 30$ au) and high eccentricities 
($e \gtrsim 0.97$). The initial nodal longitude and initial perihelion argument of orbits in the ecliptic
frame were chosen to be uniformly random.
Consequently, the population of scattered bodies initially appeared as a relatively 
thin disk near the ecliptic, and this -- when looked at by an observer in the ecliptic  
node\footnote{The ecliptic node is defined here as the intersection of the Galactic and ecliptic planes near 
the Galactic longitude $l=186^\circ$ (e.g., Fouchard et al. 2023).} -- formed the main axis of the Oort spiral in Fig. \ref{spiral}.

\begin{figure}
\epsscale{0.8}
%gr.faherty3.f and gr.spiral_orbits.f in Cluster
%\plotone{scatter_edited2.JPG}
\plotone{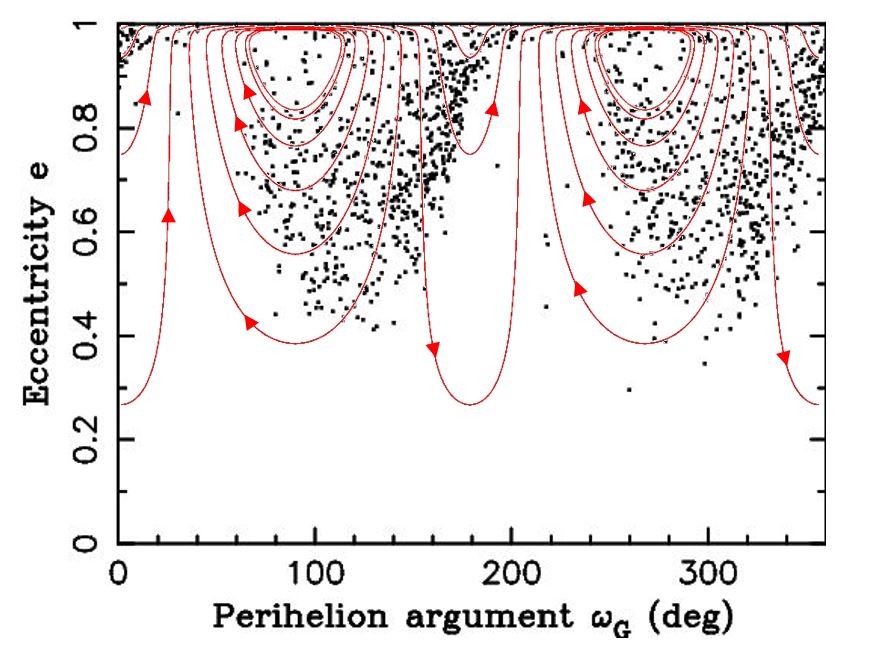}
\caption{The orbital elements of bodies in the inner Oort cloud ($a \sim 3,\!000$ au). 
  Most orbits in the inner Oort cloud are expected to have $\omega_{\rm G}=70$-180$^\circ$ or
  $\omega_{\rm G}=250$-360$^\circ$. These two broad concentrations in $\omega_{\rm G}$ appear as two spiral arms
  in Figs. \ref{spiral} and \ref{spiral2}.
  The red curves are the evolutionary tracks of $e$ and $\omega_{\rm G}$ computed from the analytic model for $a=3,\!000$ au
  and $J_z=0.070534$ (see Section~3).}            
\label{scatter1}
\end{figure}

The Galactic tide is important for orbits with $a>1,\!000$ au. As the ecliptic plane is inclined 
$\simeq 60.2$ deg with respect to the Galactic plane, bodies scattered near the ecliptic to $a>1,\!000$ au 
had large orbital inclinations in the Galactic frame ($i_{\rm G} \sim 60^\circ$), and were subject to Kozai 
cycles (Kozai 1962; we discuss the Kozai cycles in detail in Section 3). The Kozai cycles produce
anti-correlated oscillations of $e$ and $i_{\rm G}$ that are accompanied by a slow evolution of 
perihelion argument $\omega_{\rm G}$ (Fig. \ref{scatter1}). The fate of an orbit then depends on the initial value of 
$\omega_{\rm G}$. If $0^\circ < \omega_{\rm G} < 90^\circ$ or $180^\circ < \omega_{\rm G} < 270^\circ$, the orbital 
eccentricity increases, the perihelion distance drops, and the body tends to be 
scattered by planets away from the $1,\!000<a<10,\!000$ au region (often to interstellar space). 

If, instead, $90^\circ < \omega_{\rm G} < 180^\circ$ or $270^\circ < \omega_{\rm G} < 360^\circ$,
the orbital eccentricity initially decreases, and the orbit can decouple from planetary perturbations 
(Fig. \ref{scatter1}). These orbits can be stable over long timescales. They stay in the inner Oort cloud and 
continue to evolve by Kozai cycles. As $e$ decreases in a Kozai cycle, $i_{\rm G}$ increases 
from the initial value of $i_{\rm G} \sim 60^\circ$, and the orbit becomes nearly polar in the Galactic 
frame. This is when the orbital changes due to Galactic tide become exceedingly slow and orbits freeze
(Section 3). This explains the twisted arms of the Oort spiral in Figure \ref{spiral} that reach away from 
the ecliptic plane toward the Galactic poles.      

\begin{figure}
\epsscale{0.8}
%gr.faherty3.f in Cluster
\plotone{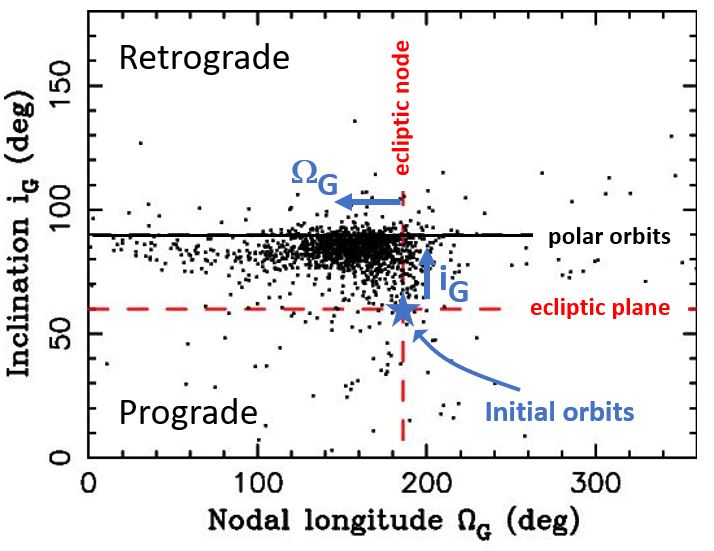}
\caption{The orbital elements of bodies in the inner Oort cloud ($a \sim 3,\!000$ au). The figure shows 
that bodies in the inner Oort cloud are expected to have nearly polar orbits in the Galactic frame 
($i_{\rm G}=75$-90$^\circ$) and orbital planes that are only slightly rotated away from the 
ecliptic ($\Omega_{\rm G}=120$-180$^\circ$; the ascending node of the ecliptic is at the Galactic longitude
$l=186^\circ$ -- the vertical red line). This implies a disk-like structure in physical space. The blue star  
is the initial location of bodies scattered by planets near the ecliptic plane. }            
\label{scatter}
\end{figure}

The ascending node of the ecliptic on the Galactic plane is located near the Galactic longitude 
$l=186^\circ$ (and, by definition, at the Galactic latitude $b=0$). In the {\it Galaxy} simulation, 
we observe that orbits in the inner Oort cloud ($1,\!000<a<10,\!000$ au) start with the nodal 
longitude $\Omega_{\rm G} \simeq 186^\circ$. Subsequently, if $\omega_{\rm G}$ has the right value (see above)
and the orbit decouples from planetary perturbations, $\Omega_{\rm G}$ slowly rotates in the retrograde 
sense (Fig. \ref{scatter}). Thus, over time, orbits move away from the ecliptic plane and their initially nearly perfect 
alignment is (slightly) smeared. The observer looking at the structure along the ecliptic node will then 
not see the disk exactly edge on, which is the situation shown in Figure 1. For example, for 
$a \sim 3,\!000$ au, the characteristic rotation of $\Omega_{\rm G}$ over 4.6 Gyr is broadly 
centered at $\sim 30^\circ$ (Figure \ref{scatter}).

\section{Analytic model}

Here we adopt the analytic model from Breiter \& Ratajczak (2005), Breiter \& Ratajczak (2006) and
Higuchi et al. (2007). The model starts with the Galactic tidal potential from Heisler \& Tremaine (1986) and neglects
all components of the tide except the (largest) one that is perpendicular to the Galactic plane.
The gravitational potential of planets is neglected as well. This is an appropriate simplification
for $a>1,\!000$ au, where the effect of Galactic tide is more important than planets (Saillenfest et al.
2019). The corresponding Hamiltonian is averaged over the orbital period. Consequently, the semimajor
axis $a$ and the $z$ component of (scaled) angular momentum
\begin{equation}
J_z=\sqrt{1-e^2} \cos i_{\rm G}\ ,
\label{angj}
\end{equation}
where $i_{\rm G}$ is the inclination with respect to the Galactic frame, are constant.
As $J_z$ remains constant during orbital evolution, $e$ and $i_{\rm G}$ show anti-correlated oscillations 
defined by Eq.~(\ref{angj}). The simplified Hamiltonian becomes
\begin{equation}
C=\sin^2\!i_{\rm G} \left(1+{3 \over 2} e^2 - {5 \over 2} e^2 \cos 2 \omega_{\rm G}\right)\ ,
\label{ham} 
\end{equation} 
where $\omega_{\rm G}$ is the argument of perihelion in the Galactic reference system. For any values
$J_z=$ const. and $C=$ const., as defined by initial conditions, the evolution of $e$, $i_{\rm G}$ and 
$\omega_{\rm G}$ can be obtained from Eqs. (\ref{angj}) and (\ref{ham}).

The model of Breiter \& Ratajczak (2005) and Higuchi et al. (2007) gives explicit expressions,
in terms of special functions, for the time evolution of $e$, $i_{\rm G}$, $\omega_{\rm G}$ and $\Omega_{\rm G}$.
For example, the evolution of eccentricity is
\begin{equation}
e(t)=[e_{\rm min}^2+(e_{\rm max}^2-e_{\rm min}^2)\ {\rm cn}^2(\theta,k) ]^{1/2}\ ,
\end{equation} 
where $e_{\rm min}$ and $e_{\rm max}$ are the minimum and maximum values 
(both are computed from the initial conditions in Breiter \& Ratajczak 2005),
${\rm cn}(\theta,k)$ is the Jacobian elliptic function, $\theta(t)$ is a linear function of time
and depends on initial conditions, 
and the modulus $k$ is a constant obtained from the initial conditions. Once $e(t)$ is solved for, the 
inclination evolution can be obtained from
\begin{equation}
i_{\rm G}(t)=\arccos {J_z \over (1-e(t)^2)^{1/2} }\ .
\end{equation}  
The evolution of $\omega_{\rm G}(t)$ can then be computed from from Eq. (\ref{ham}). Finally, the 
evolution of the longitude of node $\Omega_{\rm G}(t)$ is 
\begin{equation}
\Omega_{\rm G}(t)=\Omega_{{\rm G},i} + A_2\, \Pi(\theta(t),\alpha^2,k) \ ,  
\end{equation}
where $\Omega_{{\rm G},i}$ is the initial value of $\Omega_{\rm G}$, $A_2$ and $\alpha$ are 
constants that can be obtained from initial conditions (Breiter \& Ratajczak 2005, Higuchi et al. 
2007), and $\Pi$ is an incomplete elliptic integral of the third kind.

We evaluated $e(t)$, $i_{\rm G}(t)$, $\omega_{\rm G}(t)$ and $\Omega_{\rm G}(t)$ from 
Breiter \& Ratajczak (2005) and used it to understand the results of the more complete simulation
discussed in Sect. 2. Before we proceed with this interpretation, note that the {\it Galaxy} simulation 
in Section 2 accounted for: the (1) gravitational scattering from migrating outer planets, (2) all three 
components of the Galactic tide, and (3) stellar encounters -- none of these effects is included in 
the analytic model. At least some of the differences between our {\it Galaxy} simulation and the
analytic model arise from the neglected effects.

\begin{figure}
\epsscale{0.7}
%higuchi2.f > higuchi2.dat, and anim.f 
\plotone{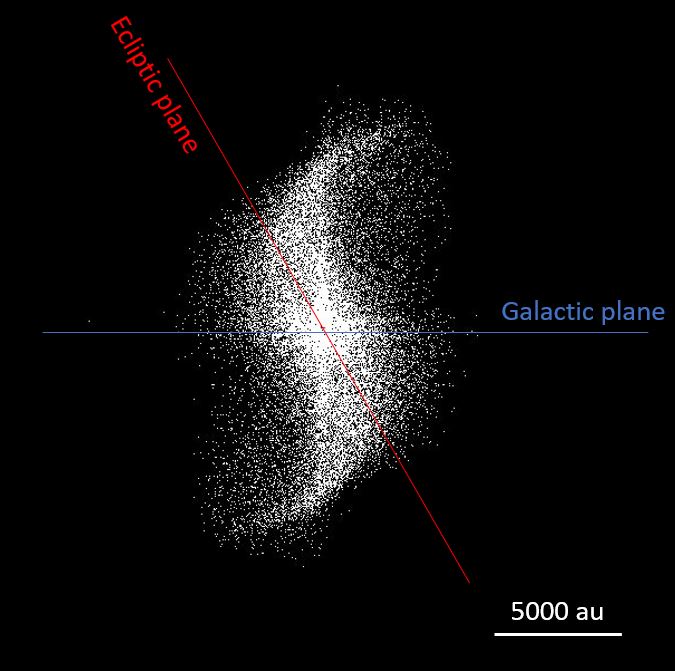}
\caption{The spiral structure reconstructed by analytical means from Breiter \& Ratajczak (2005). 
We placed 34,000 bodies onto initial orbits with $2000<a<5000$ au, $10<q<30$ au and $i<30$ deg
(inclination with respect to the ecliptic frame). The initial orbital longitudes were chosen at random.
The analytic formulas from Breiter \& Ratajczak (2005) (also see Higuchi et al. 2007) were used
to compute the orbits at $t=4.6$ Gyr, corresponding to the present epoch. The distribution of
bodies at the present epoch was then plotted in the same way as in Figure \ref{spiral}.}           
\label{higuchi2}
\end{figure}

We first checked on the results shown in Figure \ref{spiral}. For that, we generated 34,000 bodies, which 
is the same as the number of bodies shown in Fig. \ref{spiral}, and propagated their orbits with the 
analytic model for 4.6 Gyr. As a proxy for bodies scattered by planets to the inner Oort cloud distances, 
the initial orbits were given uniform distributions with $2,\!000<a<5,\!000$~au, $10<q<30$ au, $i<30^\circ$, 
and random orbital angles. The spatial distribution of bodies at $t=4.6$ Gyr, viewed from the same 
direction as in Fig. \ref{spiral}, is shown in Fig. \ref{higuchi2}. The two plots are similar in that they
show the same spiral structure. The one in Fig. \ref{higuchi2} is less fuzzy with the two arms 
standing out more clearly. There is also a concentration of orbits with $i_{\rm G} \simeq 90^\circ$
in Fig. \ref{higuchi2}, forming a vertical line that runs through the plot's center.\footnote{The
  concentration of polar orbits is probably smeared by stellar encounters in the {\it Galaxy} simulation
(Fig. \ref{spiral}).}

\begin{figure}
\epsscale{0.8}
%gr.higuchi.f
\plotone{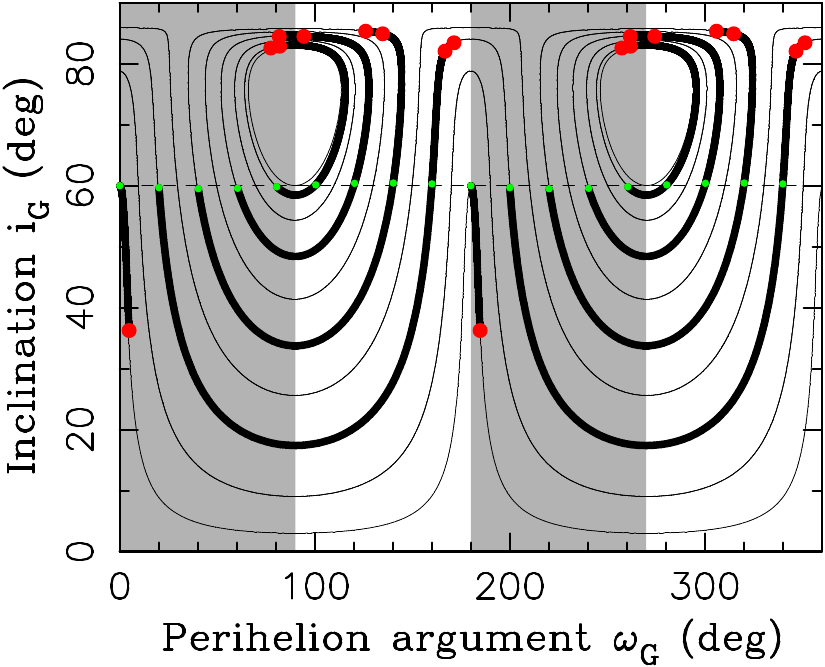}
\caption{The Galactic-tide-driven evolution of orbital inclination $i_{\rm G}$ and argument of 
perihelion $\omega_{\rm G}$, both in the Galactic reference system. The thin lines show 
trajectories for different values of the $C$ constant (Eq. \ref{ham}) and $J_z=0.070534$ (Eq. \ref{angj}). 
To  illustrate how bodies in the inner Oort cloud end up on nearly polar orbits in the Galactic frame, 
we started 18 bodies with $i_{\rm G}=60^\circ$ and $\Omega_{\rm G}=186^\circ$  (i.e., the initial orbits
in the ecliptic plane), and different values of $\omega_{\rm G}$ (equally spaced in 20$^\circ$ 
intervals; green dots on the horizontal dashed line),
and let them evolve for 4.6 Gyr (thick trajectories). The red dots label 
the final orbits. The initial orbits had $a=3,\!000$ au, $e=0.99$ and $q=30$ au.}           
\label{hig}
\end{figure}

\begin{figure}
\epsscale{0.8}
%gr.higuchi.f
%\plotone{fig6.eps}
\plotone{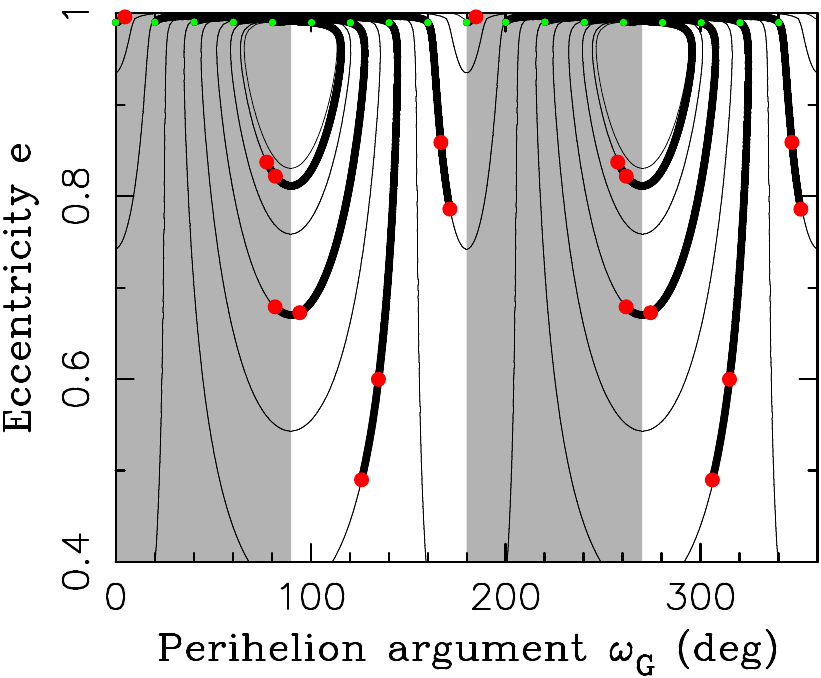}
\caption{The Galactic-tide-driven evolution of orbital eccentricity $e$ and argument of 
perihelion $\omega_{\rm G}$. The thin lines show 
trajectories for different values of the $C$ constant (Eq. \ref{ham}) and $J=0.070534$ 
(Eq. \ref{angj}). The green and red dots label the initial and final orbits, respectively. 
See Fig. \ref{hig} for additional information.}           
\label{hig2}
\end{figure}

To set up a simple analytic case, we adopted fixed $a=3000$ au, $e=0.99$ and $q=30$ au. We assumed that 
scattering from the outer planets does not produce very large orbital inclinations with respect to 
the ecliptic, as indicated by our numerical simulations ($i \lesssim 30^\circ$; Nesvorn\'y et al. 2023). 
For simplicity we therefore set $i=0$. This implies $i_{\rm G}=60.2$ deg and $\Omega_{\rm G}=186$ deg.
With these choices, the orbital evolution only depends on the initial value of $\omega_{\rm G}$ 
(Figs. \ref{hig} and \ref{hig2}). The results are closely aligned with those discussed in Sect.~2.    
If $0^\circ < \omega_{\rm G} < 90^\circ$ or $180^\circ < \omega_{\rm G} < 270^\circ$, the orbital eccentricity 
increases (Fig. \ref{hig2}), the perihelion distance drops, and the subsequent evolution would 
be affected by the scattering encounters with planets (not included in the analytic model). This 
explains why orbits with $0^\circ < \omega_{\rm G} < 90^\circ$ or $180^\circ < \omega_{\rm G} < 270^\circ$
are underrepresented in the Oort spiral. 

For $90^\circ < \omega_{\rm G} < 180^\circ$ or $270^\circ < \omega_{\rm G} < 360^\circ$, the orbital eccentricity 
initially decreases, so the orbit can decouple from planetary perturbations and become stable.  
At the same time, as $i_{\rm G}$ of the orbit increases (Fig. \ref{hig}), the orbit becomes nearly polar
and practically freezes (red dots in Fig. \ref{hig}). This is a consequence of Eq. (\ref{ham}), where 
$C \rightarrow 0$ when $i_{\rm G} \rightarrow 90^\circ$, and the orbit evolution becomes exceedingly slow.  
Two examples of this are shown in Fig. 3 in Higuchi et al. 2007. The analytic model therefore explains 
why orbits in the Oort spiral often have $i_{\rm G}=75$-90$^\circ$ (Fig. \ref{scatter}). The slow retrograde 
circulation of $\Omega_{\rm G}$ that starts near $\Omega_{\rm G}=186^\circ$ and stalls when 
$i_{\rm G} \simeq 90^\circ$ complements the picture. It explains why orbits in the inner Oort cloud often have 
$\Omega_{\rm G}=120$-180$^\circ$ (Fig. \ref{scatter}).\footnote{We also tested cases with $a<1,\!000$ au 
and $a>10,\!000$ au. For $a<1,\!000$ au, the timescales of Kozai oscillations produced by the Galactic 
tide are exceedingly long and orbits remain coupled to the outer planets. For  $a>10,\!000$ au,
the timescales of Kozai cycles are (much) shorter than the age of the Solar System (Fig.~2 in Higuchi et al. 
2007). This means that orbits can cycle up and down in $e$ and $i_{\rm G}$, and the evolution of 
$\Omega_{\rm G}$ and $\omega_{\rm G}$ is faster as well. This effectively randomizes the spatial distribution
of bodies and gives the outer Oort cloud its nearly isotropic appearance.} 

\section{Observational Detectability}

The observational detection of the Oort spiral is difficult. The reflected light from large bodies
in the inner Oort cloud can be detected by large telescopes. For example, Sheppard et al. (2019)
used the 8.2-m Subaru telescope to discover 541132 Lele\={a}k\={u}honua
with $a=1085$~au, $e=0.94$ and $i=11.7$ deg (original barycentric elements). In the Galactic reference
system, we have $i_{\rm G}=50.4$ deg, $\Omega_{\rm G}=166^\circ$ and $\omega_{\rm G}=333^\circ$,
which would allow us to project 541132 Lele\-{a}k\-{u}honua in plots like those shown in Figs. \ref{scatter1}
and \ref{scatter} (note that the semimajor axis of 541132 is roughly 2-5 times smaller than that of the bodies
shown in these figures). We infer that the orbit of 541132 Lele\={a}k\={u}honua must have had a
more complex history than simple planet scattering plus Kozai cycles. This is because 541132
Lele\={a}k\={u}honua does not have nearly as polar orbit in the Galactic frame as the bulk of inner
Oort cloud objects in Fig. \ref{scatter}. Saillenfest et al. (2019) already showed that objects
with $a \sim 1000$ au are in a transition region where the effects of Galactic and planetary
potentials combine to produce dynamical chaos.
\footnote{Another interesting object is 2021 RR205 with $a = 991$ au, $e = 0.944$ and $i=7.6$ deg.}

It may have been scattered away from the ecliptic such that the initial inclination was $i_{\rm G}<50^\circ$.
The Kozai cycles would then plausibly produce the current orbit by decreasing $e$ and increasing 
$i_{\rm G}$. Complicating factors include the effects of stellar encounters, stellar cluster and 
the possible planet 9 (Sheppard et al. 2019). In any case, the Oort spiral is mainly contributed by bodies with 
large perihelion distances and $a>2,000$ au. The telescopic observations would therefore need to detect 
objects on even more extreme orbits than 541132 Lele\={a}k\={u}honua, and obtain sufficient statistics 
for these bodies, to be able to piece together their spatial distribution. This task will have to 
wait for the next generation of telescopic surveys. 

\begin{figure}
\epsscale{0.7}
% gr.faherty2, gr.spiral bins the data... 
\plotone{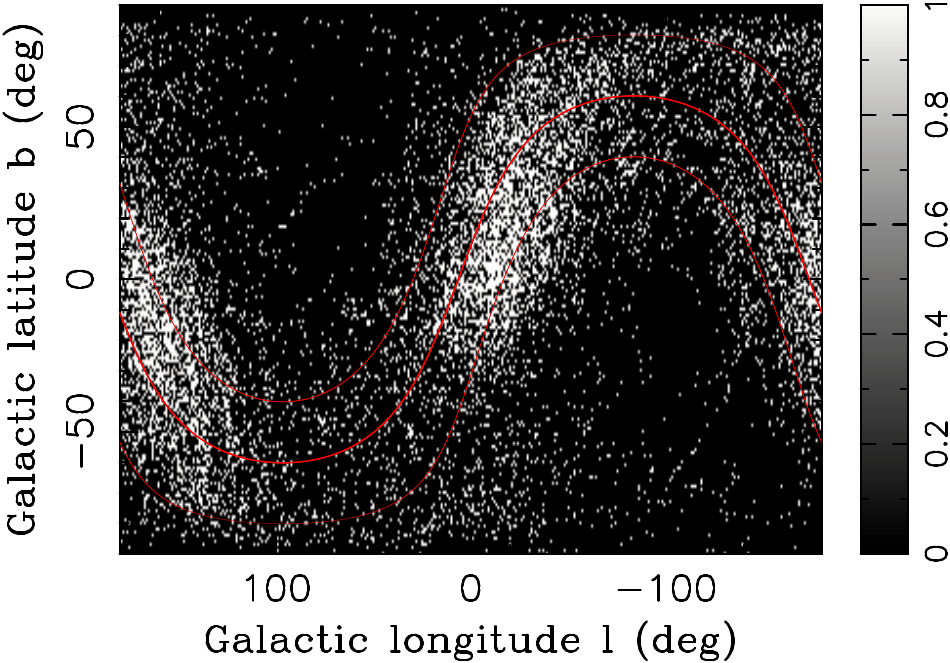}\\[5.mm]
\plotone{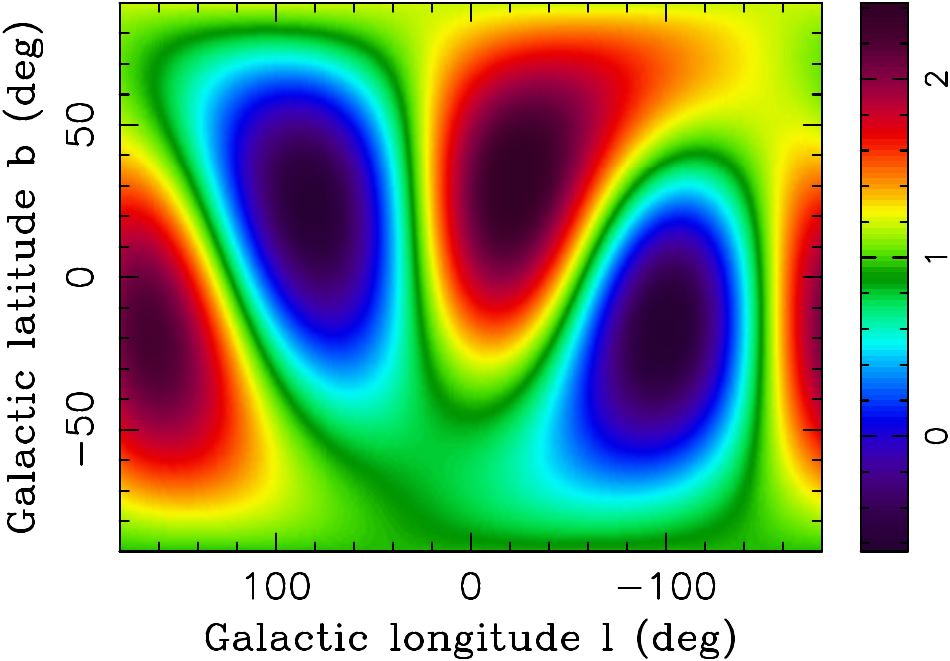}
\caption{The distribution of inner Oort cloud objects on the sky. We collected bodies with 
  $1,\!000<a<3,\!000$ au in the {\it Galaxy} simulation (Section 2) and show them here from
  the viewpoint of a terrestrial observer (upper panel). The red lines in the upper panel
  are lines of constant ecliptic latitudes $\beta=-20^\circ$, 0 and 20$^\circ$.  
  The bottom panel shows the multipole
  expansion for $l \leq 3$ with the warmer colors indicating higher number densities. The scaling 
is arbitrary.}           
\label{obs1}
\end{figure}

Figure \ref{obs1} shows the distribution of inner Oort cloud bodies on the sky as seen from
the perspective of a terrestrial observer. The two clouds in Figure \ref{obs1} correspond 
to the two spiral arms in Figs. \ref{spiral} and \ref{spiral2}.
The maximum density occurs near the Galactic coordinates $l=340^\circ$, $b=30^\circ$ and $l=160^\circ$, 
$b=-20^\circ$. We used the multipole expansion to highlight the large scale features. The 
multipole expansion of function $f$ is  
\begin{equation}
f(\theta,\phi)=\sum_{l=0}^n \sum_{m=-l}^l a_{l,m} Y_{l,m}(\theta,\phi)\ , 
\end{equation} 
where $\theta=90^\circ-b$ is the co-latitude and $\phi$ is the longitude, $a_{l,m}$ are coefficients,
and $Y_{l,m}$ are the spherical harmonics. The coefficients are computed by integration,
\begin{equation}
a_{l,m}=\int_{\cal S} f(\theta,\phi) Y^*_{l,m}(\theta,\phi)  {\rm d}\Omega \ ,
\end{equation}
over the celestial sphere ${\cal S}$, where $f(\theta,\phi)$ is the number density of objects on 
the sky, $Y^*_{l,m}$ is the complex conjugate of $Y_{l,m}$, and ${\rm d}\Omega=\sin \theta {\rm d}\theta
{\rm d}\phi$ is the infinitesimal solid angle.

\begin{figure}
\epsscale{0.7}
% gr.faherty2, gr.spiral bins the data... 
\plotone{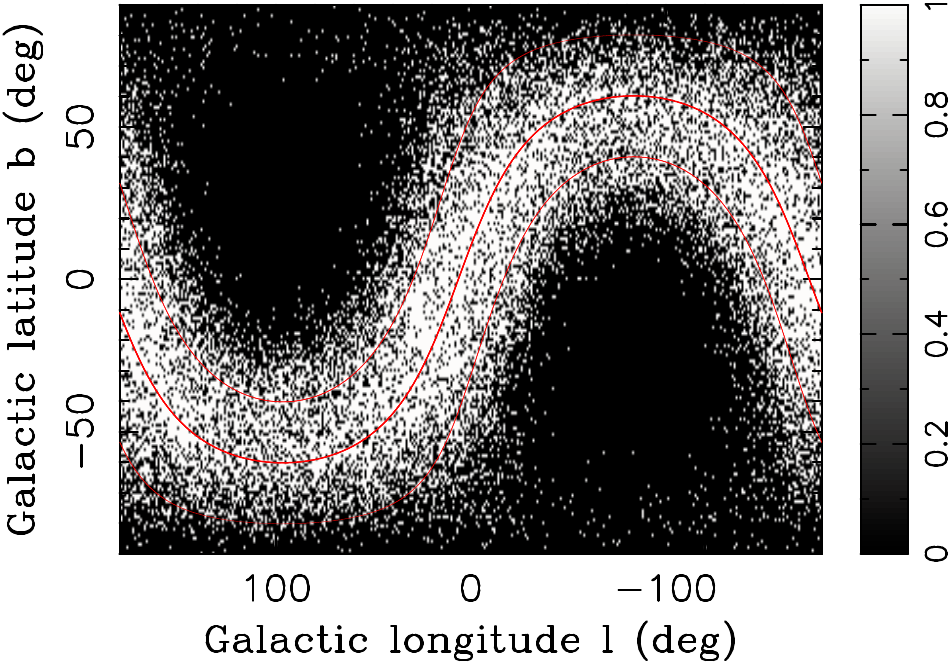}\\[5.mm]
\plotone{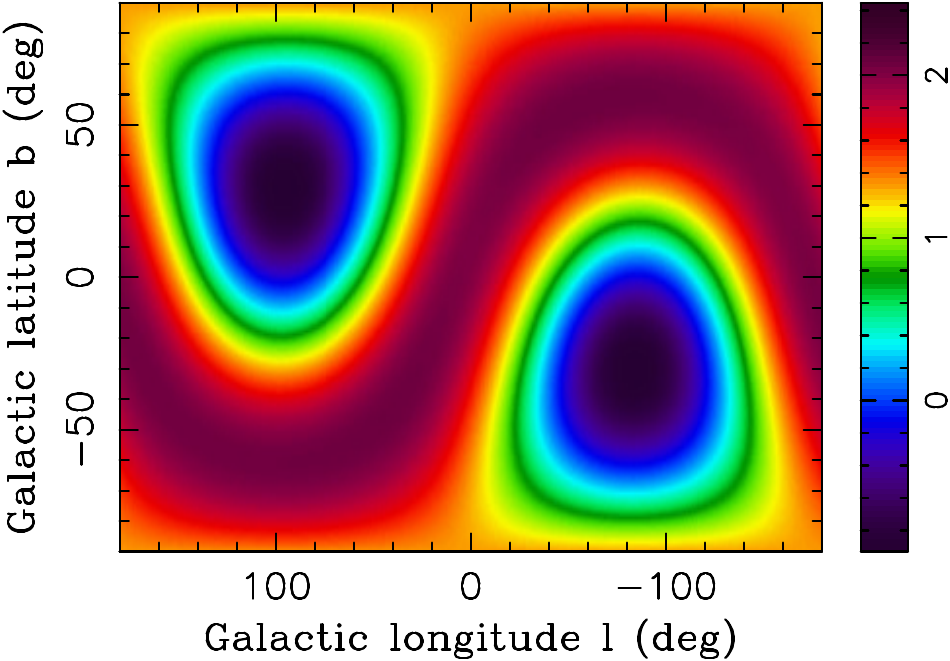}
\caption{The distribution of KBOs/SDOs on the sky. We collected bodies with 
  $30<a<1,\!000$ au in the {\it Galaxy} simulation (Section 2) and show them here from
  the viewpoint of a terrestrial observer (upper panel). The red lines in the upper panel
  are lines of constant ecliptic latitudes $\beta=-20^\circ$, 0 and 20$^\circ$.  
  The bottom panel shows the multipole expansion for $l \leq 3$ with the warmer colors
  indicating higher number densities. The scaling is arbitrary.}           
\label{obs3}
\end{figure}

\begin{figure}
\epsscale{0.7}
%gr.faherty2.f > faherty2.dat in Cluster/galaxy, and anim.f 
\plotone{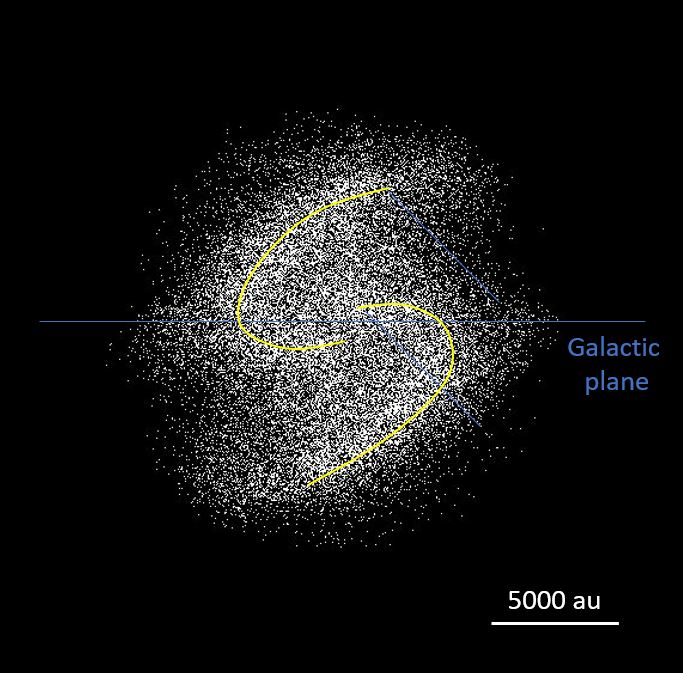}
\caption{A top view of the Oort spiral from the perspective of a distant observer at the Galactic plane. The 
  plot shows the same bodies as in Figure \ref{spiral}, but the view is rotated by $90^\circ$ around the Galactic pole.
  The yellow curves highlight the two spiral arms.}           
\label{spiral2}
\end{figure}

We find that the inner Oort cloud distribution is dominated by the quadratic term with $l=2$ and 
$m=2$ (Figure \ref{obs1}, bottom panel). For comparison, the distribution of Kuiper belt objects
(KBOs) and scattered disk objects (SDOs) with $a<1,\!000$ au is more continuous along the ecliptic
(Fig. \ref{obs3}). When represented by the multipole expansion, the distribution of KBOs/SDOs appears
to be dominated by two quadratic terms with $l=2$: $m=-1$ and $m=2$. This offers a criterion for
distinguishing the inner Oort cloud objects from KBOs/SDOs: for them the multipole expansion should be
dominated by the $a_{2,2}$ term ($|a_{2,2}| \sim 2|a_{2,-1}|$). For a more continuous distribution
such as the one shown in Fig. \ref{obs3}, $|a_{2,-1}| \gtrsim |a_{2,2}|$.   

Detecting thermal emission from small dust particles in the inner Oort cloud is similarly difficult.
For shorter wavelengths, $\lambda \lesssim 100$ $\mu$m, the large-scale thermal emission is dominated 
by the zodiacal light (e.g., Nesvorn\'y et al. 2010, Planck Collaboration 2014).
For longer wavelengths, $\lambda \gtrsim 500$ $\mu$m, the thermal emission is dominated by
the Cosmic Microwave Background (CMB) and Galactic sources (e.g., Planck Collaboration 2020).
The CMB shows an anomalous quadrupole term that is somewhat smaller than the expectations from the
best-fit cosmological model (Spergel et al. 2003, Planck Collaboration 2020). The
quadrupole moment expected from the inner Oort cloud emission has nearly the opposite orientation
on the sky (Fig. \ref{obs1}) to that of the CMB quadrupole, and would therefore -- at least in principle
-- act to decrease the CMB quadrupole. In practice, however, we estimate that the amplitude of the
inner Oort quadrupole should be some three orders below that of the CMB quadrupole ($\sim 10$ Jy/sr for Oort
vs. $\sim 10$ kJy/sr for CMB at frequencies $\nu \sim 160$ GHz), and therefore
negligible.\footnote{For this estimate, we placed $< 1$ M$_{\rm Earth}$ of material (Nesvorn\'y
  et al. 2023) between 2,000 and 5,000 au, and considered a full size spectrum of bodies between
  15 $\mu$m and 100 m -- dust grains with $a \sim 1000$ au and radii $< 15$ $\mu$m are ejected
  due to an electrical force from charging of grains outside the heliosphere (Belyaev \& Rafikov 2010).
  The size distribution was assumed to follow the equilibrium slope with the cumulative
  slope index $q \simeq 2.5$. We used the SIRT code from Nesvorn\'y et al. (2010) to estimate the thermal
  emission from this population.}  

The best chances to detect thermal emission from the inner Oort cloud are near wavelengths 
$\sim 300$-400 $\mu$m (Baxter et al. 2018). This is where the spectral radiance of both the zodiacal
and CMB emissions is reduced by a factor of $\gtrsim 10^3$ from their peak values. The small particles 
in the inner Oort cloud, $\sim 10$--100 $\mu$m, would have temperatures $T\sim10$~K and efficiently 
emit at these wavelengths. To isolate this component from the zodiacal cloud, one would have to search
for an anomalous quadrupole signature corresponding to a distribution that is not completely 
smooth in the ecliptic longitude and peaks near the expected directions (Fig. \ref{obs1}). To 
separate it from the CMB, one would have to demonstrate that the amplitude of the CMB quadrupole 
moment decreases -- relative to lower and higher multipole moments -- for these intermediate wavelengths.

The high-frequency instrument aboard the Planck satellite observed at 857 GHz,
which translates to $\lambda \simeq 350$ $\mu$m. A careful analysis of these observations could perhaps
unveil some interesting features. But even here the prospects of the inner Oort cloud detection 
are feeble. A complication arises because the thermal signal from the inner Oort particles scales with 
$1/r^2$ and temperature $T \simeq 280/\sqrt{r}$. Thus, even though these particles spend more time
near aphelion, where they contribute to the Oort spiral, they are much more easily detectable 
when they approach the terrestrial observer during their perihelion passage. This creates a strong
bias in detectability of particles with small perihelion distances; these particles have
high eccentricities and a different distribution on the sky than the one shown in Fig. \ref{obs1}.
Unfortunately, we do not have enough statistics in the {\it Galaxy} simulation to accurately predict the
expected signature from this population.

Finally, we briefly comment on Baxter et al. (2018) who used the high-frequency Planck detector (545 GHz
and 857 GHz) to search for signatures of exo-Oort clouds around nearby (hot) stars. They found several
candidates but pointed out that the rate of false positives in these observations is expected to be
relatively high. They argued that future CMB surveys and targeted observations with far-infrared and
millimeter wavelength telescopes have the potential to detect exo-Oort clouds or other extended sources
of thermal emission beyond $\sim 1\!000$ au from the parent stars. Here we point out that resolving
a spiral-like signature such as the one shown in Fig. \ref{spiral} or Fig. \ref{spiral2} would 
potentially represent a more definitive evidence for the existence of the (inner) exo-Oort cloud around
a nearby star. For the exo-Oort spiral to form, there needs to be a planetary system capable of
ejecting small bodies to $\sim 1$,000-10,000 au from its host star, and the orbital plane of
the ejected bodies needs to be significantly inclined to the Galactic plane for the Kozai cycles
to happen. 

\section{Conclusions}

We used numerical simulations and analytical methods to demonstrate that the {\it inner} Oort cloud is a
warped disk with a spiral structure roughly 15,000 au in length (Figs. \ref{spiral} and \ref{spiral2}). The
spiral forms when small bodies are scattered by planets to 1,000-10,000 au and orbitally evolve by the
Galactic tide. At 1,000-10,000 au, where the dynamical timescales are comparable to the age of the solar
system, the Galactic tide acts to: (a) raise perihelion distances and decouple bodies from planetary
perturbations, (b) rotate orbital planes such that they evolve to become nearly perpendicular to the
Galactic plane (and $\Omega_{\rm G}=120$-180$^\circ$; Fig. \ref{scatter}), and (c) favor long term stability
of orbits with $90^\circ < \omega_{\rm G} < 180^\circ$ and $270^\circ < \omega_{\rm G} < 360^\circ$. Item
(b) implies that the inner Oort cloud would look like a disk to a distant observer. Item (c)
implies two preferred directions for orbits distributed in the disk. As the preferred directions depend
on the period of Kozai oscillations (Section 3 and Higuchi et al. 2007), which in turn depends on
the semimajor axis, the inner Oort cloud bodies appear to be concentrated in two spiral arms
(Fig. \ref{spiral2}).

In retrospect, the existence of the Oort cloud spiral could have been inferred from Fig. 3 in Fouchard
et al. (2018), where the non-uniformity of orbital angles corresponds to the spiral structure reported
here. The structure was not present in Fig. 1 of Fouchard et al. (2018), which corresponds to an
initially isotropic Oort cloud, showing that the spiral structure is indeed linked to the initial
state of the Oort cloud (i.e., orbital scattering of bodies from the outer planet region along the
ecliptic). Aditionally, Fouchard et al. (2018, 2023) identified a related group of LPCs (B3) with
properties that would reflect the structure of the inner Oort cloud -- the so-called `empty
ecliptic' feature -- this feature was identified in LPC catalogs (Fouchard et al. 2023). In this sense,
the Oort cloud spiral has (indirectly) been detected.

Direct observational detection of the Oort spiral is difficult. Either (i) this structure can be pieced together
from detection of a large number of objects with $a>1,000$ au and $q>30$ au, or (ii) the thermal emission
from small particles in the Oort spiral will be separated from various foreground and background sources.
As for (ii), perhaps the best chance is to scrutinize emission at intermediate wavelengths ($\lambda \sim
100$-500 $\mu$m; e.g., Planck's 857 GHz detector) and demonstrate excess emission that is not fully
continuous in ecliptic longitude (i.e., not from the zodiacal cloud) and deviates in some significant
way from the low-degree multipoles observed at longer wavelengths (i.e., not a CMB quadrupole). Observational
detection of Oort spirals around Milky Way stars is similarly challenging (Baxter et al. 2018).

\section{Appendix A: Dynamical simulations}   

Here we make use of the dynamical model from Nesvorn\'y et al. (2023). To start with we 
disregard their cases with the stellar cluster and focus on the simulation called {\it Galaxy}.
This simulation included: the (1) migration model for the outer planets, (2) effects of planet 
scattering on disk planetesimals, and (3) Galactic potential and stellar encounters. The 
model results were calibrated on Dark Energy Survey (DES) detections of objects in the trans-Neptunian 
region (Bernardinelli et al. 2022). Here is a brief description of these components:  

{\it (1) Migration model.} The numerical simulations consisted of tracking the orbits of the 
four giant planets (Jupiter to Neptune) and a large number of planetesimals. Uranus and Neptune were 
initially placed inside of their current orbits, and were migrated outward. The {\tt swift\_rmvs4} 
code, part of the {\it Swift} $N$-body integration package (Levison \& Duncan 1994), was used to follow 
all orbits. The code was modified to include artificial forces that mimic the radial migration and 
damping of planetary orbits. The migration history of planets was informed by the best models of 
planetary migration/instability. Specifically, they adopted the migration model s10/30j from Nesvorn\'y 
et al. (2020) that worked well to satisfy many constraints. See that work for a detailed description 
of the migration parameters (e.g., migration $e$-fold timescale $\tau=10$ Myr for $t<10$ Myr and instability 
at $t=10$ Myr). The migration model also accounted for the jitter that Neptune's orbit experienced 
due to close encounters with massive bodies (Nesvorn\'y \& Vokrouhlick\'y 2016). 

{\it (2) Planetesimal Disk.} The simulations included one million disk planetesimals distributed from 
4 au to beyond 30 au. Such a high resolution was needed to obtain good statistics for the Oort cloud. 
The initial surface density of disk planetesimals was assumed to follow the truncated power-law profile 
from Nesvorn\'y et al. 2020 (also see Gomes et al. 2004). The step in the surface density at 30 au was parameterized 
by the contrast parameter $c \sim 10^3$, which is simply the ratio of surface densities on either side 
of 30 au (the planetesimals with initial $a>30$ au are not an important source for the Oort cloud). 
The initial eccentricities and inclinations of orbits were set according to the Rayleigh 
distribution with scale parameters $\sigma_e=0.1$ and $\sigma_i=0.05$. The disk bodies were assumed 
to be massless such that their gravity did not interfere with the migration/damping routines. 

{\it (3) Galactic potential and stellar encounters.} The Galaxy was assumed to be axisymmetric and the 
Sun followed a circular orbit in the Galactic midplane (Sun's migration in the Galaxy was not included; 
Kaib et al. 2011). The Galactic tidal acceleration was taken from Heisler \& Tremaine (1986) 
(see also Wiegert \& Tremaine 1999, Levison et al. 2001). The stellar mass density in the solar neighborhood was 
set to $\rho_0=0.15$ $M_\odot$ pc$^{-3}$. The simulations accounted for the effect of stellar encounters. 
The stellar mass and number density of different stellar species were computed from Heisler et al. 
(1987). The stars were released and removed at the heliocentric distance of 1 pc (206,000 au).  For each 
species, the velocity distribution was approximated by the isotropic Maxwell--Boltzmann distribution. 
The dynamical effect of passing molecular clouds was ignored.

{\it (4) Calibration on observations.} The {\it Galaxy} simulation was run over 4.6 Gyr at which point 
the orbital distribution of bodies in the trans-Neptunian region was 
compared with DES observations (Bernardinelli et al. 2022).
DES covered a contiguous 5000 $\deg^2$ of the southern sky between 2013-2019, with the majority of the 
imaged area being at high ecliptic latitudes. The search for outer Solar System objects yielded 812 
Kuiper belt objects (KBOs) with well-characterized discovery biases, including over 200 SDOs with $a > 50$~au.
The DES survey simulator\footnote{The simulator enables comparisons between population models and the 
DES data by simulating the discoverability conditions of each member of the test population. It is  
publicly available on \texttt{GitHub} - \url{https://github.com/bernardinelli/DESTNOSIM}} 
(Bernardinelli et al. 2022) was used to bias the model in the same way as the data. Nesvorn\'y et al. 
(2023) made use of DES observations to calibrate the magnitude distribution of trans-Neptunian objects 
and establish that the dynamical model results were consistent with DES observations. The model 
fidelity was previously tested from observations of short-period comets (Nesvorn\'y et al. 2017) and 
LPCs (Vokrouhlick\'y et al. 2019).
 
\acknowledgements

\begin{center}
{\bf Acknowledgments} 
\end{center} 
\vspace*{-3.mm}
The {\it Galaxy} simulation was performed on the NASA Pleiades Supercomputer. We thank the NASA
NAS computing division for continued support. The work of DN was funded by the NASA
EW program. DV acknowledges support from the grant 25-16507S of the Czech Science
Foundation.
%We thank an anonymous reviewer for critical comments.

\clearpage

\clearpage

\clearpage

\clearpage

\clearpage

\clearpage

%\clearpage
%\begin{figure}
%\epsscale{0.48}
% gr.fagerty2.d in Dones, pgfswift sirt_particle, sirt_particle > test2.dat, 
% expansion > test3.dat, gr.expansion, evince a.ps
%\plotone{expand1.eps}
%\plotone{expand2.eps}
%\plotone{expand3.eps}
%\plotone{expand4.eps}
%\caption{Multipole expansion of the inner Oort cloud emission in the 350 $\mu$m wavelengths ($l \leq5 $). 
%In (a), the emission is contributed by grains with heliocentric distances $r>1000$ au. In (b),
%the emission is contributed by grains with heliocentric distances $r>1500$ au. These results were
%obtained from the dones simulation, from particles with $1000<a<2000$ au, $q>30$ au and with 100 Myr
%sampling. For some reason, the spiral structure in dones simulations appears to be closer, at 
%1000-2000 au, then in the galaxy simulation. This could happen because I used different matter
%densities in the stellar neighborhood (0.2 vs. 0.15). There is 0.3 Earth mass and particles as 
%small as 10 um and as large as 100 km. Simple radial profile of emission temperature.}          
%\label{expand}
%\end{figure}

\clearpage

\end{document}